\documentclass[aps,pre,twocolumn,showpacs]{revtex4}
\usepackage{graphics,epsfig}
\usepackage{epsfig,graphicx}
\usepackage{dcolumn}
\usepackage{bm}
\usepackage{times}

\begin{document}

\title{Evolutionary prisoner's dilemma game with dynamic preferential selection}

\author{Zhi-Xi Wu$^{1}$, Xin-Jian Xu$^{2}$, Zi-Gang Huang$^{1}$, Sheng-Jun Wang$^{1}$, and Ying-Hai Wang$^{1}$}

\address{$^{1}$Institute of Theoretical Physics, Lanzhou
University, Lanzhou Gansu 730000, China\\
$^{2}$Department of Electronic Engineering, City University of
Hong Kong, Kowloon, Hong Kong SAR, China}

\date{\today}

\begin{abstract}
We study a modified prisoner's dilemma game taking place on
two-dimensional disordered square lattices. The players are pure
strategists and can either cooperate or defect with their
immediate neighbors. In the generations each player update its
strategy by following one of the neighboring strategies with a
probability dependent on the payoff difference. The neighbor
selection obeys a dynamic preferential rule, i.e., the more
frequently a neighbor's strategy was adopted by the focal player
in the previous rounds, the larger probability it will be chosen
to refer to in the subsequent rounds. It is found that cooperation
is substantially promoted due to this simple selection mechanism.
Corresponding analysis is provided by the investigations of the
distribution of players' impact weights, persistence, and as well
as correlation function.
\end{abstract}

\pacs{02.50.Le, 05.50.+q, 87.23.Cc, 89.65.-s}

\maketitle

\section{introduction}
Game theory and its evolutionary context are efficient frameworks
to study complex behaviors of biological, ecological, social and
economic systems \cite{Neumann,Maynard,Axelrod,Hofbauer}. Of
particular renown is the evolutionary prisoner's dilemma game
(PDG) which has attracted much attention in theoretical and
experimental studies \cite{Axelrod,Hofbauer,Wahl,Fehr,Mesterton}.
In the original PDG, players can make two choices: either to
cooperate with their co-players or to defect. They are offered
some payoffs dependent on their choices, which can be expressed by
a $2\times2$ payoff matrix in agreement with the four
possibilities. The players get rewards $R$ (or punishment $P$) if
both choose to cooperate (or defect). If one player cooperates
while the other defects, then the cooperator $(C)$ gets the lowest
payoff $S$ (sucker's payoff), while the defector $(D)$ gains the
highest payoff $T$ (the temptation to defect). Thus the elements
of the payoff matrix satisfy the conditions: $T>R>P>S$ and
$2R>T+S$, so that lead to a so-called dilemma situation where
mutual cooperation is beneficial in a long perspective but egoism
can produce big short-term profits.

One of the most interesting items on the PDG is to study under
what conditions mutual cooperation will emerge and sustain stably
or how to facilitate the cooperation of the whole population
\cite{Maynard, Axelrod, Hofbauer}. In the evolutionary PDG, the
state that all players are defectors has been proved to be an
evolutionary stable state \cite{Hofbauer}, which has inspired
numerous investigations of suitable extensions that enable
cooperation to persist. Nowak and May \cite{Nowak} introduced a
spatial evolutionary PDG model in which individuals located on a
lattice play with their neighbors and with themselves. The
dynamics is governed by a deterministic rule: in each subsequent
round, individuals adopt the strategy that has gained the highest
payoff among its neighbors (including also themselves) in the
previous round. It has been shown that the spatial effects promote
substantially the survival of cooperators \cite{Nowak, Hubermann,
Nowak_1, Nowak_2, Hauert_0}. The enhancement of cooperation on
lattice-like spatial structure is robust even if some specified
distance interactions are allowed \cite{Nowak_2}.

Szab\'{o} and T\H{o}ke extended the deterministic dynamics to a
stochastic evolutionary one \cite{Szabo_0}: rather than following
the most successful neighbor's strategy straightly, the adoption
of one of the neighboring strategies is allowed with a probability
dependent on the payoff difference. This revised version took into
account the irrational choices of the players. It was observed
that a stable absorbing state of all $C$ emerged when the
temptation to defect is below a certain critical value
(noise-dependent). Vainstein and Arenzon studied the PDG on a
diluted lattice \cite{Vainstein} and found that cooperation is
easier to maintain due to blocking the spreading of defection
caused by the empty sites on the lattice. The investigations of
the PDG on random graphs have shown that cooperation is strongly
enhanced for lattices with fluctuating connectivity in comparison
with fixed connectivity lattice \cite{Duran}. In the past few
years, the PDG has been studied on different social network
models. It was found that cooperation can be maintained on these
networks in a wide range of network parameters \cite{Abramson,
Kim, Ebel, Holme, Wu, Szabo_1, Szabo_2}. In addition, several
mechanisms were also introduced to sustain high concentration of
cooperators, such as voluntary participation \cite{Szabo_1},
dynamic network model \cite{Zimmermann} and dynamic payoff
matrices \cite{Tomochi}, and so on. Very recently, Santos \emph{et
al.} studied the PDG and snowdrift games on scale-free networks.
Their results indicated that the heterogeneity of networks favors
the emergence of cooperation \cite{Santos}.

Though many intriguing fruits have been obtained in understanding
the emergence of cooperation in spatial PDG \cite {Nowak,
Hubermann, Nowak_1, Nowak_2, Hauert_0, Szabo_0, Vainstein, Duran,
Abramson, Kim, Ebel, Holme, Wu, Szabo_1, Szabo_2}, it would also
be interesting to explore other mechanisms to enhance the
cooperative behavior. In the present work, we make further studies
of the evolutionary PDG using Szab\'{o}-T\H{o}ke version
\cite{Szabo_0} on two-dimensional disordered square lattices.
Considering individuals heterogeneously affected by their
neighbors in society, we introduce impact weight to each one of
the interacting players. Based on the player's impact weights, a
dynamic preferential selection (DPS) mechanism is incorporated
into the spatial PDG to model dynamic behaviors of human
communities. It will be shown that the DPS promotes substantially
the cooperative behavior of the players which may provide a new
perspective in understanding the persistence of cooperation in
realistic world.

In the next section, evolutionary rules of the game are explained.
In Sec. III, simulation results of our model implemented on
disordered lattices are presented and, in Sec. IV, the effects of
the DPS on the evolution of cooperation is studied in detail.
Conclusions are given in the last section.

\section{The Model}
We consider the evolutionary PDG with players located on the
two-dimensional disordered square lattice which is characterized
by a total $\phi$ portion of randomly rewired links with fixed
number of neighbors of each site (in the limit
$\phi\rightarrow1.0$, the \lq\lq disordered lattice\rq\rq is in
fact a regular random graph \cite{Szabo_1, Szabo_3}). This
approach excludes those effects coming from the fluctuation of the
number of neighbors as it happens on diluted lattices \cite
{Vainstein}, random graphs \cite{Duran}, and scale-free networks
\cite{Santos}. Similar consideration has also been made in Refs.
\cite{Szabo_2, Szabo_3}. The players are pure strategists and can
follow only two simple strategies: $C$ (cooperate) and $D$
(defect). Each player plays a PDG with itself and with its
neighbors, and collects payoffs dependent on the payoff-matrix
parameters. The total payoffs of a certain player is the sum over
all interactions. Following common practices \cite{Nowak, Szabo_0,
Abramson, Kim}, the elements of the payoff matrix can be rescaled,
i.e., choosing $R=1$, $P=S=0$, and $T=b$ $(1.0<b<2.0)$  to
represent the advantage of defectors over cooperators. We have
checked that the properties of the simulation results do not
change for $S=-\epsilon$ $(0.0 < \epsilon \ll 1.0)$.

In society, some special persons may influence others much
stronger than the average individual \cite{Kim}. In other words,
different neighbors would have different impacts on one's
behavior. In general, one can expect that the influence between
two individuals would be asymmetric and time-dependent. To model
this situation, we define a quantity $A_{ij}(t)$, which describes
the impact weight of the $j$th player on the $i$th player at time
$t$ and possesses the asymmetric property, i.e., $A_{ij}(t) \neq
A_{ji}(t)$. In this way, we hope to catch some general effects of
the dynamic asymmetric influence among interacting players on the
dynamical behavior of the game.

After each round of the game, players are allowed to inspect their
neighbors' payoffs and change their strategies in the next round
according to the comparisons. The randomly chosen player $i$
revises its strategy by selecting one of its neighbors $j$ with a
probability $\gamma$ in terms of a preferential selection rule:
\begin {equation}
\gamma_{ij}=\frac{A_{ij}(t)}{\sum_{k\in\Omega_{i}} A_{ik}(t)}\label{rule1},
\end {equation}
where $\Omega_{i}$ is the community composing of the nearest
neighbors of $i$. Eq. (\ref {rule1}) means that the stronger the
impact of a neighbor is, the larger probability it is selected to
compare with. If and only if their strategies are different, the
$i$th player's state as well as the neighbor's impact weight will
be updated, otherwise nothing happens (no strategy transformation
and impact weight update). Following ideas developed by Szab\'{o}
\emph{et al.} \cite{Szabo_0, Szabo_1, Szabo_2, Szabo_3}, given the
total payoffs ($E_i$ and $E_j$ for the $i$th and $j$th players,
respectively) from the previous round, the player $i$ adopts $j$'s
strategy with the probability
\begin{equation}
W = \frac{1}{1 + \exp{[-(E_j - E_i)/\kappa]}}\label{eq1},
\end{equation}
where $\kappa$ characterizes the noise introduced to permit
irrational choices. It should be noted that the transformation is
only affected by their payoff difference. We set the value of
$\kappa$ to $0.1$, similar to that in Refs. \cite{Szabo_0,
Szabo_1, Szabo_2, Szabo_3}. Generating a random number $r$ from a
uniform distribution in the interval $(0, 1)$, if $r< W$ (or
$r>W$), the player $j$'s strategy is accepted (or discarded) by
the player $i$ and $A_{ij}(t)$ is revised according to the
following rule
\begin {equation}
A_{ij}(t+1)=A_{ij}(t)(1\pm\alpha)\label{rule2},
\end {equation}
where the minus corresponds to the case of $r>W$, denoting no
strategy updating for the $i$th player. Initially, all $A_{ij}(0)$
are assigned to $1.0$. The parameter $\alpha$ in Eq. (\ref{rule2})
can be depicted as a multiplication factor which characterizes
qualitatively the relative change of the impact weight in each
round. Larger $\alpha$ corresponds to stronger strengthening (or
losing) of impact. In subsequent investigations, the value of
$\alpha$ is constrained in the range $(0,1)$ to ensure no negative
elements in $A_{ij}(t)$, which means that all opinions ($C$ and
$D$ in the present case) have opportunities to be followed. If
strategies of the players spread successfully, they can be called
as \lq\lq winners\rq\rq regardless of the payoffs (for example,
due to the noise, the worse performance strategy also has a
certain probability to take over a better performance one). Thus,
this rule (Eq. (\ref{rule2})) could be termed as: \lq\lq
win-strengthen, lose-weaken\rq\rq. The more frequently a player's
strategy followed by its neighbors, the larger probability it will
be picked up to refer to in the subsequent generations, and vice
versa.

\section{simulation results}
In what follows three groups of systems will be considered: (i)
regular square lattices ($\phi=0.0$ with $\alpha=0.0$, $0.01$,
$0.1$, and $0.3$, respectively); (ii) weak disordered lattices
($\phi=0.1$ with the same values of $\alpha$ as before); (iii)
strong disordered lattices ($\phi=1.0$ with the same values of
$\alpha$ as the former). On the one hand (as a byproduct), we want
to understand how the underlying disordered structures affect the
evolution of the PDG; on the other hand (as the main aim), we try
to explore the potentiality of the DPS (in combination with the
disordered structures) on the evolution of cooperation.

\begin{figure}
\centerline{\epsfxsize=7cm \epsffile{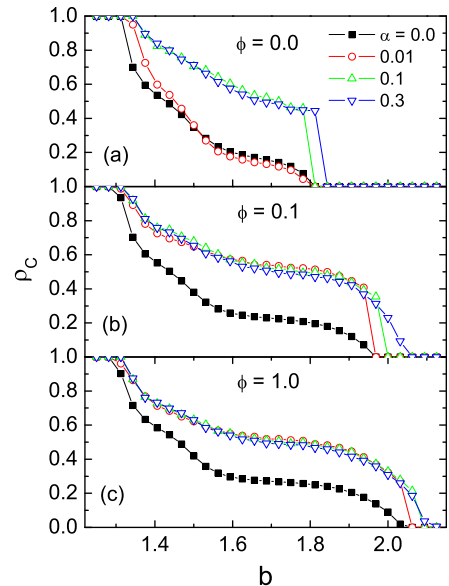}} \caption{(color
online). Average density of cooperators, $\rho_{c}$, as a function
of the temptation to defect $b$ in the equilibrium state. Filled
and open symbols correspond to the cases of random selection and
preferential selection, respectively.} \label {fig1}
\end{figure}

\begin{figure}
\centerline{\epsfxsize=7cm \epsffile{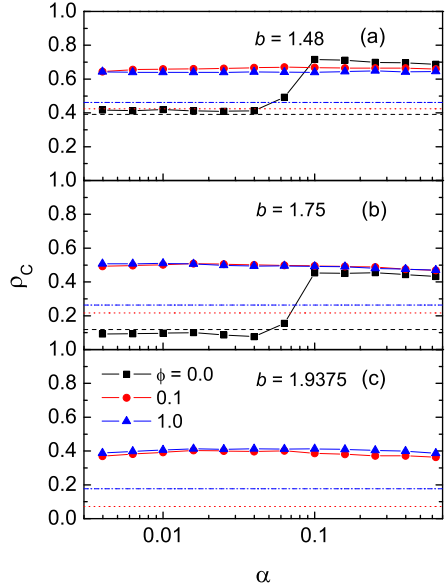}} \caption{(color
online). Average density of cooperators, $\rho_{c}$, as a function
of the multiplication factor, $\alpha$, for three special values
of $b$ (colored symbols). The colored lines correspond to random
selection cases $(\alpha=0.0)$: dashed (black) for $\phi=0.0$,
dotted (red) for $\phi=0.1$, and dash-dotted (blue) for
$\phi=1.0$, respectively.} \label {fig2}
\end{figure}

All the simulations are performed in systems with $300\times300$
players. Starting from a random initial state with equal fraction
of $C$ and $D$, we iterate the model with synchronous update. The
evolution eventually leads to dynamic equilibrium states with
small fluctuations of the cooperator density around an average
value. The parameter $\alpha$, which determines the evolution of
impact weights, is not critical, and convergence has been verified
for all values of $\alpha$ between $0$ and $1$. A key quantity is
the density of cooperators, $\rho_C$, which is defined as the
average fraction of players adopting the strategy $C$ in the
equilibrium state. The total sampling times were $10^5$
Monte-Carlo (MC) steps \cite{Note_1}, and the equilibrium density
of cooperators was obtained by averaging over the last $10^4$
steps. Each data point (in Figs. \ref{fig1} and \ref{fig2})
results from an average over 10 realizations for the same type of
disordered lattices specified by the parameter $\phi$.

The dependence of $\rho_C$ on $b$ in the equilibrium state for
different values of $\phi$ and $\alpha$, are shown in Fig.
\ref{fig1}. The main features in the steady state are similar to
the results reported in Ref. \cite{Szabo_0}, i.e., there exist two
different absorbing states $(\rho_C=1$ and $ \rho_C=0)$ separated
by the active phase (coexistence of $C$ and $D$). We have found
numerically that $\rho_C=1$ in all the cases we are interested in
when $b<5/4$, which can be regarded as a homogeneous cooperation
state. Since our main aim goes beyond this trivial steady-state,
we will only concentrate on the region of $b>5/4$, where new
features may emerge.

First, we consider the model without DPS ($\alpha=0.0$). In the
case of $\phi=0.0$, which corresponds to the regular square
lattice, we recover the results of the stochastic model
\cite{Szabo_0}, i.e. $\rho_C$ decreases monotonically with the
increase of $b$ and vanishes at a threshold $b_c$. When long range
links emerge on the lattice ($\phi=0.1$), the level of cooperation
is promoted (a larger threshold $b_c$ in Fig. \ref{fig1}(b) than
that in Fig. \ref{fig1}(a)), which is different from previous
reports that local interactions promote the sustainment of
cooperation (see Refs. \cite{Nowak, Hubermann, Nowak_1, Hauert_0}
and the references therein). Particularly, in the case of
$\phi=1.0$ (corresponding to a regular random graph), where the
spatial correlation is very weak, cooperation is enhanced further
(if it is just measured by the value of $b_c$). Even in the case
of $b>2.0$, minor fractional cooperators can be found in a sea of
defectors.

These results can be understood in the following way. On the one
hand, in our model, the state transformation of the players
conforms to a Fermi function (see Eq. (\ref{eq1})) suggested by
Szab\'{o} \emph{et al.} \cite{Szabo_0, Szabo_1, Szabo_2, Szabo_3}.
Using this update rule, they have also found that long range
interactions enhance the cooperation when the PDG (with no
self-interactions included) was performed on regular random graphs
\cite{Szabo_1}. On the other hand, self-interactions are included
in the present work. Recently, Hauert and Doebeli have studied
another famous evolutionary game, snowdrift game, on different
types of lattice \cite{Hauert_1}. They have found that the spatial
structure eliminates cooperation for intermediate and high
cost-to-benefit ratio of cooperation because benefits of costly
cooperative acts accrue not only to others but also to the
cooperator itself \cite{Hauert_1, Doebeli}. As to our PDG model,
each player $C$ plays with itself besides its nearest neighbors,
which indicates that it will gain at least $R$ payoff even in the
worst case (surrounding by four defectors). In a different
interpretation, besides their neighbors, the cooperators'
investment will benefit themselves too. Note that the high
cost-to-benefit ratio of cooperation in snowdrift game corresponds
to large values of $b$ in the PDG. Considering these two factors,
it is not surprising that the disordered structure can promote
cooperation in the present model.

We now focus our attention on the influence of the DPS on the game
evolution. The results obtained for $\alpha=0.01, 0.1, 0.3$ are
summarized in Fig. \ref{fig1}. Although qualitative behaviors are
similar to those of the random selection case, there are some
remarkable quantitative differences. The simulation results depend
strongly on both the parameters $\phi$ and $\alpha$. For well
structured populations ($\phi=0$, regular square lattice case)
with $\alpha=0.01$, the DPS promotes cooperation for small $b$ in
comparison with the random selection case. For large $b$, however,
the tendency is reversed and the fraction of cooperators is
somewhat lower than the corresponding value in the case of
$\alpha=0.0$. There exists a cross point for $b$ separating the
two regions which depends on $\alpha$: for $\alpha=0.01$,
$b\approx1.5$. For the other two values of $\alpha$ ($0.1$ and
$0.3$), the DPS enhances greatly the cooperation with respect to
otherwise. In the cases of disordered structural populations
($\phi=0.1$ and $1.0$), for all the values of $\alpha$ considered
here, the cooperative behavior can be substantially promoted (see
Figs. \ref{fig1}(b) and \ref{fig1}(c)) and maintained even in the
extreme defection circumstance ($b>2.0$), where the game is not a
proper Prisoner's Dilemma. It is strikingly interesting that, for
disordered structural populations, all the curves nearly collapse
into a single curve (shown in Figs. \ref{fig1}(b) and
\ref{fig1}(c)) when the DPS is introduced into the game. The
visible difference exists only in the region where cooperators are
going to extinction. The larger $\alpha$ gives rise to larger
threshold $b_c$. The data collapsing is also found in well
structured populations for large $\alpha$ (see Fig.
\ref{fig1}(a)).

From these observed features, we argue that, whether the
underlying \lq\lq lattice\rq\rq is well structured or disordered,
as soon as the multiplication factor $\alpha$ is finite and
sufficiently large, the enhancement of cooperation will be to some
extent realized. Despite of the fact that the disordered structure
prefers to favor the emergence of cooperation, the much stronger
enhancement of cooperative behavior for $\alpha>0.0$ than that for
$\alpha=0.0$ implies that the DPS plays a crucial role in the
dynamics.

\begin{figure}
\centerline{\epsfxsize=9cm \epsffile{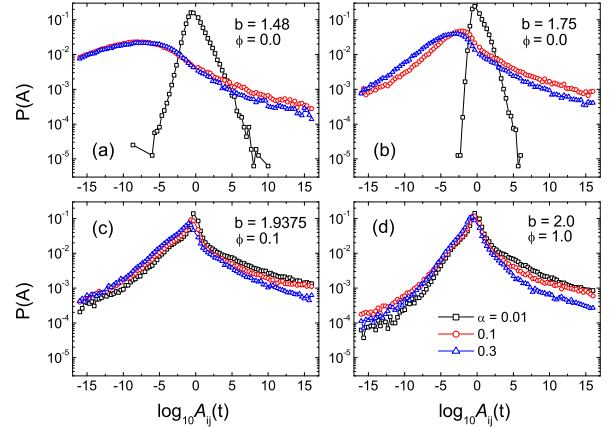}} \caption{(color
online). Histograms (over $10^5$ time steps) of impact weights of
players in the equilibrium state. Only the data which fall into
the region $[10^{-16}, 10^{16}]$ are shown. The power-law decaying
behavior determined by the intrinsic nature of Eq. (\ref{rule2})
is visible in both the very large limit and the very small limit
of the impact weights.} \label {fig3}
\end{figure}

In Fig. \ref{fig2}, we show the plots of $\rho_C$ as a function of
$\alpha$ for three special values of $b$, to emphasize the changes
in behaviors as the lattice varies. The colored dashed lines
correspond to the results obtained for the random selection case
and the symbols for the DPS case. Different colors denote
different disordered degree of the lattice: black for $\phi=0.0$,
red for $\phi=0.1$ and blue for $\phi=1.0$, respectively. Note
that only the curves for $\phi=0.0$ have a clear sensitivity to
the values of $\alpha$ (black squares in Figs. \ref{fig2}(a) and
\ref{fig2}(b)). The temptation to defect, $b$, has also an
influence on the results. In the cases of small $\alpha$ with
$\phi=0.0$, for low temptation ($b=1.48$), the cooperation level
is slightly higher for the DPS case than that for the random
selection case; whereas for high temptation ($b=1.75$), the
phenomenon is reversed. The Similar discrepancy can also be found
for large values of $\alpha$ with $\phi=0.0$: the spatial
structure promotes slightly the cooperation level for $b=1.48$
(see Fig. \ref{fig2}(a)), and yet inhibits that for $b=1.75$ (see
Fig. \ref{fig2}(b)). The transition point seems to be around
$\alpha \approx 0.06$. Since regular lattices are not realistic
representations of most actual population structures
\cite{Newman}, especially human (because of the mobility and
dispersal ability of the individuals), we do not comment further
on these trivial results and focus our attention on those issues
where disordered structures are present ($\phi=0.1$ and $1.0$).

Now taking a look at Figs. \ref{fig2}(a), (b) and (c), we clearly
observe that for any finite values of $\alpha$ we are interested
in, given fixed values of $b$ and $\phi$, the cooperation level is
almost independent of $\alpha$, and the relative deviation of the
results is less than $5\%$. Despite of this point, we would like
to point out that the enhancement of cooperation appears also
insensitive to the special values of $\phi$, i.e., more disordered
structures have little essential influence on the evolutionary
results, at least, for the case of $0.1<\phi<1.0$. This problem
should deserve to make further investigations in the future work.

To end up this section, let us review the novel features of the
results obtained by computer simulations. When introducing the DPS
to the evolutionary PDG, we have found that the cooperative
behavior in the populations is substantially enhanced with respect
to the random selection case. The enhancement is robust against
whether the underlying lattice is well structured or disordered.
Though the evolutionary results depend on both $\phi$ and
$\alpha$, as long as the underlying lattice is finitely disordered
and the multiplication factor $\alpha$ is also sufficiently large
(e.g., $\alpha>0.003$ in the present studied cases \cite{Note_2}),
the resulting cooperation level seems insensitive to both $\phi$
and $\alpha$ for fixed values of $b$.

\begin{figure*}
\centerline{\epsfxsize=15cm \epsffile{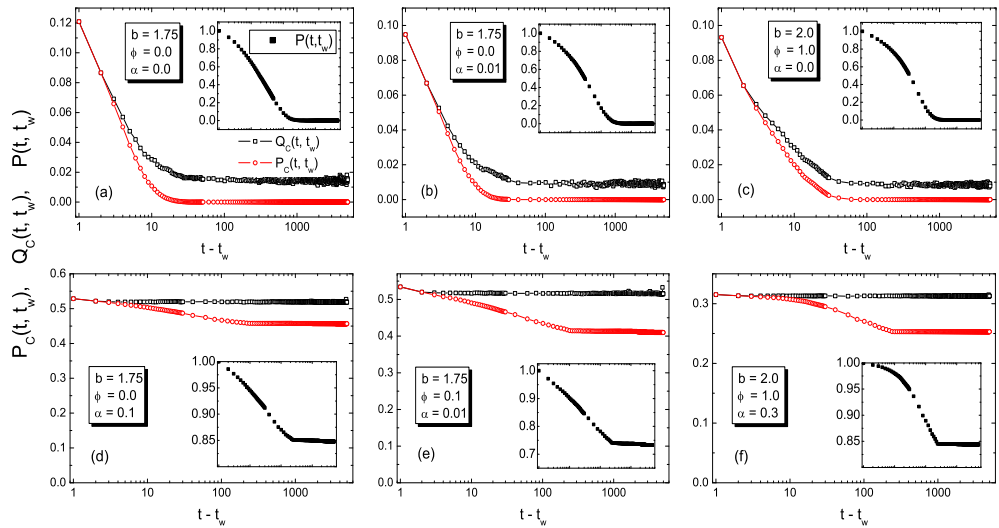}} \caption{(color
online). Persistence function $P_{c}(t,t_w)$ and correlation
function $Q_{c}(t,t_w)$ for different combinations of $b$, $\phi$
and $\alpha$ as a function of time interval $t-t_w$ where
$t_w=95000$. Inset: time evolution of the persistence function
$P(t,t_w)$ for both strategies $C$ and $D$. That the persistence
$P_{c}(t,t_w)$ goes to zero in the long time limit indicates the
random walk and annihilation of cooperators, whereas the non-zero
plateau of it implies the stable maintenance of the communities of
the cooperators. The plots of $Q_{c}(t,t_w)$ also give the same
hints, which sustains the level of the datum time $t_w$ for the
stable maintenance case, and displays the mean-field behavior for
the random walk case. See the text for details.} \label {fig4}
\end{figure*}

\section{analysis and discussion}
Before discussing the further details as to why the above
evolutionary results may occur, we briefly review those results
obtained for the evolutionary PDG performed on the regular
lattice. Generally, we are more interested in the region near the
extinction threshold of cooperators, from which we can figure out
clearly cooperators' evolution and surviving in the sea of
defectors. For this issue, Hauert \textit{et al.} \cite{Szabo_1,
Hauert_1} have found that cooperators can survive by forming
large, compact clusters which minimize the exploitation by
defectors. Along the boundary, cooperators can outweigh their
losses against defectors by gains from interactions within the
cluster . Similar phenomena were observed earlier by Szab\'{o} and
T\H{o}ke \cite{Szabo_0}. The compact clusters composing of
cooperators do random walks and spread out in the background of
defectors. Though, occasionally a cluster splits into two or two
clusters collide, merge, or annihilate and vanish \cite{Szabo_1},
the compact structures of the surviving clusters can remain in
principle unchanged.

The problem can be viewed more explicitly: when the players'
strategy update is implemented by randomly selecting a neighbor to
compare with according to Eq. (\ref{rule2}), the compact
structures favor those internal cooperators to forming a stable
core, since the components interact only with players taking over
strategy $C$. The existence of this core enables those cooperators
on surface to have enough channels to collect payoffs so that they
can resist the invading of defectors, which in return reinforce
the stability of the core. The mutual protection (or reciprocity)
of the cooperators enables the compact structures to sustain
stably, and hence contributes to the persistence of cooperation.

Motivated by the above interpretations, some conjecture on the
origin of those results of our model with the presence of DPS
depicted in Sec. III may be appropriate here. We think that the
DPS could induce the emergence of pairs of influential co-players,
and when some of them are cooperators, they will \lq\lq
always\rq\rq refer to their influential neighbors' strategies in
the subsequent generations, then communities consisting of their
neighbors and themselves may form and survive stably in the
background of defectors, which would contribute to the persistence
of cooperation. Those influential co-players taking over strategy
$C$ can be regarded as the core of the communities (or clusters).
To test this speculation, we check the probability distribution of
$A_{ij}(t)$ in the equilibrium state, which is expected to have a
broad distribution so that there could arise a large enough
fraction of influential co-players (being cooperators) to form
stable core in populations.

\begin{figure*}
\centerline{\epsfxsize=13cm \epsffile{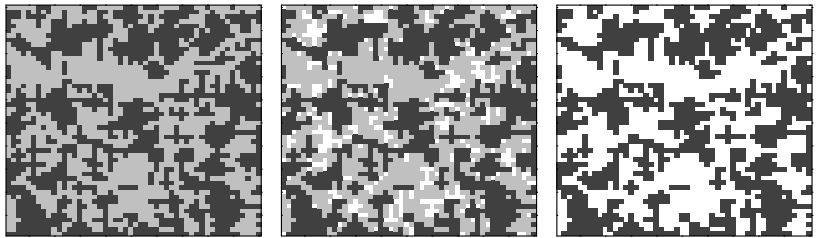}} \caption{Right
panel: illustration of cooperators and defectors at time $t=9.5
\times 10^4$. Middle panel: snapshot showing players that remain
unchanged between MC time $t=9.5 \times 10^4$ and $10^5$: the
\emph{pinned} cooperators, defectors, and the \emph{active}
players are denoted by dark gray boxes, light gray boxes, and
white vacancy, respectively. Right panel: only the \emph{pinned}
cooperators are shown. Parameters: $\phi=0.0, b=1.75,
\alpha=0.1$.} \label{fig5}
\end{figure*}

In Figure \ref{fig3}, we plot probability distributions of
$A_{ij}(t)$ for four different combinations of $b$ and $\phi$ and
with corresponding three values of $\alpha$: $b=1.48$, $\phi=0.0$
(Fig. \ref{fig3}(a)); $b=1.75$, $\phi=0.0$ (Fig. \ref{fig3}(b));
$b=1.9375$, $\phi=0.1$ (Fig. \ref{fig3}(c)); and $b=2.0$,
$\phi=1.0$ (Fig. \ref{fig3}(d)), respectively. As indicated, there
are very large broad distributions of $A_{ij}(t)$ except for two
combinations, ($b=1.48$, $\phi=0.0$, $\alpha=0.01$) and ($b=1.75$,
$\phi=0.0$, $\alpha=0.01$). These two combinations do not favor
the emergence of cooperation (see Fig. \ref{fig1}(a) for the
symbols of red circles) in contrast to other combinations, and
also do not provide sufficient conditions for players to form
stable clusters which can be seen in the following part. It is
worthy to point out that the combinations of well structured
populations with sufficiently large $\alpha$ values (Figs.
\ref{fig3}(a) and \ref{fig3}(b)) or disordered populations with
finite $\alpha$ values (Figs. \ref{fig3}(c) and \ref{fig3}(d))
resulting in broad histograms of $A_{ij}(t)$ indeed facilitate the
cooperative behavior compared with corresponding random selection
cases $(\alpha=0.0)$, which have been verified by previous
simulations (see Figs. \ref{fig1} and \ref{fig2}).

To prove that some fractional players may form stable clusters in
the presence of the DPS in a more detailed way, we measured the
persistence function $P_{C}(t,t_w)$, the fraction of cooperators
that do not change strategy between an initial waiting time $t_w$,
and the time $t\geq t_w$ \cite{Vainstein}. In addition, the
correlation function $Q_{C}(t,t_w)$, which characterizes the
fraction of cooperators at time $t$ that have arisen at time $t_w$
in spite of finitely many times flipping of the strategy during
the two time interval, is also explored. In a distinct view, the
results of these two quantities as a function of the time interval
$t-t_w$ (measured in MC steps, the time $t_w$ can be selected
arbitrarily as long as the systems had attained equilibrium) are
summarized in Fig. \ref{fig4} on a log-linear scale for
$t_w=95000$.

For the cases of well structured populations with sufficiently
large $\alpha$ (Fig. \ref{fig4}(d)) or disordered populations with
finite $\alpha$ (Figs. \ref{fig4}(e) and \ref{fig4}(f)), after an
initial decrease, the persistence attains, for large time
intervals, a plateau whose value depends on the parameters $b$,
$\phi$ and $\gamma$. Interestingly, for a non-Hamiltonian model we
studied here, the decaying behavior of the persistence in the
equilibrium is exponential at large time intervals, just like that
has been observed commonly in Hamiltonian models and in site
diluted lattice model \cite{Note_3}. If the persistence does not
go to zero, there is a fraction of sites of cooperators that flip
only finitely many times (blocking), and domain wall movements are
constrained (pinning) \cite{Vainstein}. For the present systems we
studied, it indicates that communities of cooperators exist stably
in the background of defectors. In the cases of random selection
(no matter whether the populations are well structured or
disordered, Figs. \ref{fig4}(a) and \ref{fig4}(c)) or well
structured populations with very small $\alpha$ (Fig.
\ref{fig4}(b)), however, the persistence goes to zero in the long
time interval limit, which means that all the cooperators are
renewed completely after finite waiting time. This is reminiscent
of the random walk and annihilation \cite{Szabo_0}. The
persistence function $P(t,t_w)$ for both strategies $C$ and $D$
shown in insets also displays the similar characteristics.

The time dependence of the correlation function $Q_{C}(t,t_w)$
also reflects the same evolving behavior of the systems. For well
structured populations with sufficiently large $\alpha$ (Fig.
\ref{fig4}(d)) or disordered populations with finite $\alpha$
(Figs. \ref{fig4}(e) and \ref{fig4}(f)), $Q_{C}(t,t_w)$ fluctuates
weakly around the average fraction of cooperators, $\rho_C$,
indicating the stable maintenance of the communities of
cooperators. For the random selection case (Figs. \ref{fig4}(a)
and \ref{fig4}(c)) or well structured populations with very small
$\alpha$ (Fig. \ref{fig4}(b)), the random walks and annihilation
of cooperators causes the long time correlation to be independent
of the initial state, which can be calculated roughly by a
mean-field method. Since the spread of cooperators can be regarded
as walking randomly in such cases, the probability of revisiting
those sites which had been visited before will be proportional to
the average density of cooperators. Assuming that the fraction of
cooperators is equal to $\rho(t_w)$ at time $t_w$ (the system has
already attained equilibrium before this time), the mean-field
approximation gives the relation, $Q_{C}(t,t_w)\approx
\rho_C\times \rho(t_w)$, in the long time limit,
$t\rightarrow\infty$. As shown in Figs. \ref{fig4}(a),
\ref{fig4}(b) and \ref{fig4}(c), the analysis is in well agreement
with numerical simulations.

For an intuitive understanding of the surviving of cooperators, in
Fig. \ref{fig5} we present snapshots of the system in the
equilibrium state under the condition of $\phi=0, b=1.75,
\alpha=0.1$, showing both \emph{active} and \emph{pinned} players.
These snapshots are a $50 \times 50$ portion of the full $300
\times 300$ lattice. Cooperators and defectors are shown in dark
gray and light gray boxes, respectively. The white vacancy denotes
\emph{active} players. This typical configuration is almost stable
and the small amount of \emph{active} sites is confined to a few
regions. It is shown that most cooperator clusters remain
unchanged confirming previous analysis.

At last, it should be pointed out that it is only qualitatively
depicted that the large broad distributions of the impact weights
favor the formation of stable clusters. The quantitative analysis
on this issue, however, is unimplemented. The work along this line
is in progress.

\section{Conclusions}
In this work, we have explored the general question of cooperation
formation and sustainment from the perspective of the coevolution
between the dynamics of players' states and their
inter-influential relationships. By considering asymmetric and
heterogenous influential effects in many natural populations, we
defined impact weights for any pairs of neighboring individuals,
which describes the influence of one player on another and evolves
timely. Based on this quantity, a dynamic preferential selection
(DPS) mechanism was introduced to the dynamics of an evolutionary
PDG. To simulate the mobility and dispersal ability of
individuals, the disordered structure of populations was also
taken into account.

Although disordered structure has been proved to favor the
emergence of cooperation, it was found that the DPS plays a
crucial role in determining the evolutionary results of the game.
In fact, for well structured populations with strong
multiplication effects (large $\alpha$) or disordered structural
populations with the presence of the DPS, the cooperative behavior
can be substantially promoted. These findings are presented for
some specific sets of parameter values of $b$, $\phi$ and
$\alpha$, but they are qualitatively the same for a broad range of
values. By analyzing the probability distribution of $A_{ij}(t)$
in the long time limit, the persistence function $P_{C}(t,t_w)$
and the correlation function $Q_{C}(t,t_w)$ for the cooperators in
the equilibrium state, we found that the DPS gives rise to very
large broad distributions of the impact weights, which favor the
influential cooperators to form stable communities, and thereby
contribute to the emergence and persistence of cooperation. One
interesting result is that the enhancement of cooperation, once
realized, is insensitive to some special values of $\phi$ and
$\alpha$. That is to say, the enhancement of cooperation in the
presence of the DPS is general and robust, especially for finitely
disordered structural populations. Another intriguing result is
that, given the appropriate multiplication factor $\alpha$,
cooperators can always persist in the whole range of $1.0<b<2.0$
for disordered structural populations, providing an escape hatch
out of states of mutual defection predicted by both classical and
evolutionary game theory. Thus, the DPS is capable of changing the
doomed fate of cooperators and ensuring them to persist even under
harsh conditions when $b\rightarrow2.0$. Although it is simple,
the present model offers an efficient way to simulate real social
behaviors, and might shed lights in understanding the evolution of
cooperation in complex social systems where \emph{asymmetry},
\emph{heterogeneity}, and \emph{feedback} are ubiquitous.

This work was supported by the Fundamental Research Fund for
Physics and Mathematics of Lanzhou University under Grant No.
Lzu05008.


\begin{references}
\bibitem{Note_1}
For $\alpha>0.003$, $10^5$ MC steps are sufficient for the systems
to attain equilibrium. For smaller $\alpha$, a much longer
relaxation time is needed.
\bibitem{Note_2}
We have also simulated the PDG for the cases of $\alpha<0.003$.
The enhancement of cooperation and convergence have also been
verified. However, the MC time needed for the systems to attain
equilibrium increases dramatically.
\end{references}
\end{document}